%% file: main.tex
\documentclass[runningheads]{llncs}
\usepackage[T1]{fontenc}

\usepackage{graphicx}
\usepackage[table,xcdraw, dvipsnames]{xcolor}
\usepackage{listings}
\usepackage{verbatim}
\usepackage{lstautogobble}

\usepackage{bussproofs}
\usepackage{amsmath}
\usepackage{amssymb}
\usepackage{bbm}
\usepackage{stmaryrd}
\usepackage{slashbox,pict2e}
\usepackage{courier}
\usepackage{algorithm}
\usepackage{algpseudocode}
\usepackage[most]{tcolorbox}
\usepackage{todonotes}
\definecolor{dgreen}{rgb}{0,0.6,0}
\definecolor{codepurple}{rgb}{0.58,0,0.82}
\definecolor{mauve}{rgb}{0.58,0,0.82}
\newcommand\comm[1]{}
\newcommand\url[1]{#1}

\begin{document}

%
\title{A tool assisted methodology to harden programs against multi-faults injections}
\titlerunning{Hardening programs againt multi-faults injection}


\author{{Etienne Boespflug   \and Abderrahmane Bouguern
\and  Laurent Mounier  \and Marie-Laure Potet}}

\institute{{VERIMAG\\ University of Grenoble Alpes (UGA), Grenoble, France}
\email{Firstname.Lastname@univ-grenoble-alpes.fr}}


\authorrunning{{E. Boespflug \and A. Bouguern \and  L. Mounier  \and M-L Potet}}

\maketitle

\begin{abstract}

Fault attacks consist in changing the program behavior by injecting faults at run-time
in order to  break some expected  security properties.
Applications are hardened against fault attack adding countermeasures.
According to the state of the art, applications must now be protected against 
\emph{multi-fault injection} \cite{wistp/KimQ07,fdtc/TrichinaK10}. 
As a consequence developing applications which are robust becomes a
very challenging task, in particular because countermeasures can be also the target of attacks \cite{inter_cesti20,MartinKP22}.
The aim of this paper is to propose an assisted methodology for developers allowing to harden an application against
multi-fault attacks, addressing several aspects:
how to identify which parts of the code should be protected and
how to choose the most appropriate countermeasures, making the application more robust and
avoiding  useless runtime checks?


\end{abstract}

%

%
%

\section{Introduction}

\input{new-intro.tex}

%
%

\section{Fault injection attacks}
\label{sec:motivating-example}

\input{example-and-lazart.tex}

%
%

\section{Robustness metrics for  multi-faults}
\label{sec:metrics-analysis}

\input{robustness-def}


\section{Countermeasures analysis}
\label{sec:cm_analysis}

\input{cm_analysis}


\section{Our methodology for hardening program}
\label{sec:algo}

\input{algorithms}

\input{experimentations}
\label{exps}

%
%

	
%
%

\input{conclusion}

\section*{Acknowledgment}
This work is supported by SECURIOT-2-AAP FUI 23 and by the French National Research Agency in the framework of the "Investissements d’avenir” program (ANR-15-IDEX-02) and the project ARSENE of PEPR Cybersecurité.



%
\bibliographystyle{IEEEtran}
\bibliography{biblio}


\appendix
\section{Frama-C schemes}
\label{annexe1}
\input{annexeA.tex}

\end{document}

%% file: new-intro.tex
Fault injection is a powerful attack vector, allowing to modify the code and/or data of a software,
going much beyond more traditional intruder models relying ``only'' on code 
vulnerabilities and/or existing side channels to break some expected security properties. 
This technique initially targets security critical embedded systems, using physical disturbances 
(e.g., laser rays, or electro-magnetic fields) to inject faults. However, it may now also concern 
much larger software classes when considering recent hardware weaknesses like the so-called Rowhammer attack  
\cite{Seaborn15,GrussMM16}, or by exploiting some weaknesses in the power management 
modules~\cite{Tang17,QiuWLQ19,Murdock2019plundervolt}.
Furthermore, in the growing domain of IoT,
security is based on very sensitive operations such as boot-loading or
Over-the-Air firmware update which must be protected against fault injections\cite{Timmers16,app12010417}.

As a result, programs must be hardened against fault injection, combining hardware and/or
software \emph{countermeasures} aiming to  detect runtime security violations.
According to the state of the art, applications must now be protected against (spacial or temporal)
\emph{multi-fault injection} \cite{wistp/KimQ07,fdtc/TrichinaK10}, namely when several faults can be injected at 
various times or locations during an execution. As a consequence developing applications which are robust 
becomes a 
very challenging task and a cat-and-mouse game, in particular because countermeasures can be also the target of attacks
\cite{inter_cesti20,MartinKP22}.
The aim of this paper is to propose an assisted methodology for developers allowing to harden an application against
multi-fault attacks, addressing several aspects: 
how to identify which parts of the code should be protected and
how to choose the most appropriate protection schemes, making the application more robust and  
avoiding  useless runtime checks?

There exist tools adding countermeasures, generally at compilation-time. They are dedicated to particular fault models
(data modification, instruction skip, flow integrity) \cite{Hillebold2014,ProyHBC17,BellevilleHCBRSC18}, for instance by adding redundant checks or
duplicating idempotent instructions. Nevertheless these 
tools target single fault robustness, where countermeasures themselves cannot be faulted. Furthermore such countermeasures are added in a
brute force approach, based on (coarse-grained) directives given by the developers,
indicating which parts of the code must be protected. 
Such an approach is no longer realistic when multi-faults must be taken into account, since more complex countermeasures must be  
considered, potentially adding unnecessary performance/size  costs.

The objective of this paper is to address the issue of  assisting security developers in the countermeasure 
insertion process.  In particular, we provide the following contributions:  
\begin{enumerate}
  \item we formulate the problem of  robustness comparison in presence of multi-faults;
  \item we propose a methodology to analyze countermeasures ``in isolation'', both in terms of  classes of attacks 
    they detect and considering their own attack surface;
  \item we propose an algorithm allowing to harden applications starting from identified  vulnerabilities, depending on countermeasure properties;
  \item we provide an implementation of this approach, based on the Lazart tool~\cite{Potet/ICST14} and evaluate our approach on a benchmark of code examples. 
\end{enumerate}

Section \ref{sec:motivating-example} introduces  multiple fault-injection and countermeasures
 through a motivating example and  presents the Lazart tool.
Section \ref{sec:metrics-analysis} proposes definitions for robustness evaluation and 
comparison dedicated to multi-faults.
Section \ref{sec:cm_analysis} describes the proposed methodology for analyzing countermeasures in term of 
their adequacy and weakness against established fault models. 
Section \ref{sec:algo} presents our countermeasure placement algorithms and illustrates them 
on several examples.
Finally, Section \ref{sec:conclusion} discuss related works, limitations of our solution and future directions.

%% file: example-and-lazart.tex
In this section we  introduce fault injection attacks through a simple example and the tool Lazart which is used
to analyze robustness against fault attacks. Then we propose a hardened version of our example illustrating classical countermeasures
and introduce our contributions.
 
\subsection{A fragile byteArrayCompare  version}
Listing~\ref{lst:bac1} is an excerpt of a {\tt byteArrayCompare} function used in the verifyPIN collection taken from the FISSC public benchmark\cite{DureuilPPLCC16}. Such a function compares  two PIN codes and returns true if they are equal.
{\tt{BOOL\_TRUE}} and  {\tt {BOOL\_FALSE}} are robust encoding of boolean constants true and false\footnote{Classically
{\tt 0xAA} and {\tt 0x55} to be resistant to bit flip.}.
The security property  we want to guarantee is that {\tt result} is true if only if the two PIN codes are equals ({\tt a1[0..size-1]}=={\tt a2[0..size-1]}.

\lstset{caption={Fragile byteArrayCompare function},language=C,label=lst:bac1,numbers=left,xleftmargin=2em,basicstyle=\ttfamily\scriptsize}
\begin{lstlisting}
BOOL byteArrayCompare(UBYTE* a1, UBYTE* a2, UBYTE size){
      int i;
      int result = BOOL_TRUE;
      for(i = 0; i < size; i++) 
            if(a1[i] != a2[i]) result = BOOL_FALSE; 
      return result; }
\end{lstlisting}

This example is insecure against fault injection consisting in inverting control flow condition.
To show that we use 
  Lazart~\cite{Potet/ICST14},   a tool analyzing the robustness of a software under multi-fault injections.
    It takes as input a C code, a fault model,  a security property and exhaustively find all
attack paths. Lazart  relies on a 2-steps approach:
    \begin{itemize}
    \item First, a \emph{high order mutant} is generated from the LLVM representation of the program. This mutant statically encodes all the possible injected faults (as symbolic boolean values) according to a chosen fault model 
    \item Then, a \emph{dynamic-symbolic exploration}, performed by Klee~\cite{CadarDE08}, produces a representative path for
all the successful attacks violating the security property.
    \end{itemize}

Various fault models are supported by Lazart targeting control flow or data modifications (as illustrated in
Section \ref{sec:cm_analysis}). 
Table \ref{table:exemple2} line V1 gives the results supplied by Lazart for this first version
when {\tt size} is fixed to 4  assuming all bytes of the two input arrays
{\tt a1} and {\tt a2} are different. We consider here a limit of 8 faults.
All these attacks violate the security property.

\vspace{-0.5cm}

\begin{table}
\begin{scriptsize}
\begin{center}
\caption{Lazart results for the {\tt byteArrayCompare} versions}
\label{table:exemple2}
    \begin{tabular}{|c|c|c|c|c|c|c|c|c|c|c|}
        \hline
        Order & 0 fault & 1 fault & 2 faults & 3 faults & 4 faults & 5 faults & 6 faults & 7 faults & 8 faults\\
        \hline
        V1 & 0 & {\bf 1} & {\bf 1} & {\bf 1}  & {\bf 2} & 0 & 0 & 0 & 0 \\
        \hline
        V2 & 0 & 0 & {\bf 1} & 0  & {\bf 1} & 0 & {\bf 1} & 0 & {\bf 2}\\
        \hline
    \end{tabular}
\end{center}
\end{scriptsize}
\end{table}

\vspace{-0.8cm}

The 1-fault attack consists in inverting the loop condition $i<size$ (line 4 of Listing~\ref{lst:bac1}),
stopping directly the array comparison.  The 2-faults, 3-faults and one of the 4-faults attack consist in
inverting the test of line 5, one, two or three times respectively, and then stopping the loop.
The last 4-faults attack consists in inverting four times the test of line 5 and exiting the loop normally.

\subsection{A robust byteArrayCompare  version}

We propose in Listing \ref{lst:bac2} a more robust version of the function byteArrayCompare, adding classical 
redundant-check countermeasures: 1) a test line 10 checking the exit value of the loop and 2)
 a systematic verification of the result of condition (lines 6 and 8).  
Calls to the function {\tt atk\_detected} stop the execution, signaling the detection of an abnormal behavior.

\lstset{caption={Example: Secure byteArrayCompare function},language=C,label=lst:bac2,numbers=left,xleftmargin=2em,basicstyle=\ttfamily\scriptsize}
\begin{lstlisting}
BOOL protected_byteArrayCompare(UBYTE* a1, UBYTE* a2, UBYTE size){
      int i;
      int result = BOOL_TRUE;
      for(i = 0; i < size; i++) {
        if(a1[i] != a2[i]) {
           if(a1[i] == a2[i]) atk_detected();
           result = BOOL_FALSE;}
        else if(a1[i] != a2[i]) atk_detected(); }
      if(i != size)  atk_detected(); 
      return result; }
\end{lstlisting}

In a multi-fault context, the attacker is able to bypass countermeasures. For instance
here she can invert the extra tests (lines 6, 8 and 9). Results supplied by Lazart for this
new version (still for the test inversion fault model) are given on Table \ref{table:exemple2}, line V2. 
Therefore, the 1-fault attacks found in the fragile function is no longer possible and it now requires two test 
inversions:  the loop condition  and its associated check line 9. 
More generally each previous attack now requires to double the number of injected faults. 
As we can see, 
added countermeasures make our function more robust, but unfortunately they 
 also come with their own attack surface (here our encoding of countermeasures is sensitive to test inversion).

\subsection{Attack information supplied by Lazart}
\label{sec:IP}

Attacks will be represented by sequences of fault occurrences consisting in pairs (injection point, fault model).
For instance the first attack of Table \ref{table:exemple2} is represented by
< IP4(TI), IP9(TI)>, where IP4 is the fault injection point of line 4 and TI denotes ``test inversion''.
From that we can compute how many times a fault injection point occurs into successful attacks. 
Table \ref{table:hot-spot2} shows results associated to Table \ref{table:exemple2}. 
In particular we can observe that attacking IP6, corresponding to an added countermeasure, 
never results in a successful attack. 
As a consequence this countermeasure can be removed, without impacting the robustness 
of our implementation (for the considered security property).
In the same way if we only consider order 2 attacks, countermeasures lines 5 and 8 are no longer useful.
Other configurations where some countermeasures can be safely removed can be found in \cite{FDTC20}.
\begin{table}[h]
\begin{scriptsize}
\begin{center}
\caption{Lazart Hotspots analysis for  Listing~\ref{lst:bac2}}
\label{table:hot-spot2}
    \begin{tabular}{|c|c|c|c|c|c|c|c|c|c|c|c|}
        \hline
        IP  line & 0 fault & 1 fault & 2 faults & 3 faults & 4 faults & 5 faults & 6 faults & 7 faults & 8 faults & Total\\
        \hline
         line 4 & 0 & 0 & 1  & 0 & 1 & 0 & 1 & 0 & 1& {\bf 4}\\
        \hline
        line 5 & 0 & 0 & 0  & 0 & 1 & 0 & 2 & 0 & 7 & {\bf 10}\\
        \hline
         line 6 & 0 & 0 & 0  & 0 & 0 & 0 & 0 & 0 & 0 & {\bf 0}\\
        \hline
         line 8 & 0 & 0 & 0  & 0 & 1 & 0 & 2 & 0 & 7 & {\bf 10}\\
        \hline
         line 9 & 0 & 0 & 1  & 0 & 1 & 0 & 1 & 0 & 1 & {\bf 4}\\
        \hline
    \end{tabular}
\end{center}
\end{scriptsize}
\end{table}


In the context of multi-faults,  some attacks can be considered as {\it redundant}. 
An attack $b$ will be consider as redundant with respect to an attack $a$ iff $a$ is a \textit{proper prefix} of $b$,
denoted by $a \leq b$, $\leq$ being a partial order relation.  

vspace*{-1em}

\begin{definition}{Minimal attacks}
\label{def:minimal}

\noindent
Let $E$ be a set of attacks. $Minimal(E)$ is the smallest subset of $E$ such that every attack $b$ in $E$
is represented by a unique attack $a$ in $Min(E)$ such that $a \leq b$.
\end{definition}

In order to give a more synthetic view Lazart can compute the subset of minimal attacks and how many
attacks they represent.  For the example of Listing~\ref{lst:bac1} no attack is redundant, but considering at first minimal attacks
happens to be useful for larger examples \cite{Lacombe21}. 
Lazart supplies also others results allowing to evaluate the coverage 
of the analysis,  such as the number of explored paths, if some timeout has been raised, etc.  
These information are useful to assess the completeness of the analysis. 

\subsection{Our objectives and contributions}

In the context of single fault, hardening a code generally follows a ``try and test'' approach: countermeasures
are added and robustness is checked through a new analysis (as such a secure implementation is now expected to 
become robust against single-fault attacks). Nevertheless this process is not worth considering
in a multi-fault context since it would introduce a very detrimental overhead. 
The aim of this paper is to propose several placement algorithms allowing to add countermeasures advisedly
targeting to avoid unnecessary code. 
To do that we propose a methodology for stating and analyzing countermeasures ``in isolation'' in order
to formally establish properties such as the {\it adequacy} of a countermeasure w.r.t. a given fault model
and its inherent {\it attack surface}.
Countermeasures we target are systematic
countermeasures detecting attacks, like test or load duplication,
and, more generally, systematic countermeasures allowing to detect errors in data or control flows. 

To the best of our knowledge, this approach is innovative.
As pointed out before, generally tools add countermeasures in a systematic way (e.g., \textit{stack canaries} are added by compilers to any 
functions satisfying some basic syntactic criteria)
without precisely taking into account which control locations should be protected w.r.t. the security properties which have to be enforced.
This situation introduces potentially serious and unnecessary overheads.
On the other hand, methods has been proposed to prove hardened programs \cite{Christofi/CE13,Rauzy/CE14,MartinKP22}, or to verify the efficiency of
a given form of countermeasure against attacker models \cite{Goubet/CARDIS15,heydemann2019formally}.
All these approaches are  developed in the context of single fault and  adapted to a particular countermeasure or fault models.
On the contrary, the approach we consider here is general, modular, able to cope with multi-fault attacks and it addresses
the problem of optimization of countermeasures placements.
As a first step we propose in the next section a framework to properly compare the robustness of several versions of a same code in a multi-faults
context.

%% file: robustness-def.tex
In the context of certification for high levels of security \cite{common-criteria-3}
applications are submitted to vulnerability analyses 
(the AVA class of Common Criteria),  conducted by experts team  (as ITSEF laboratory).    
For instance rating physical attacks for smart cards or similar devices is established by the 
JIL Hardware Attacks Subgroup \cite{JIL/20} on the based on 
a set of relevant factors: elapsed time, expertise, knowledge of the target of evaluation, access to it,
\ldots

At the level of  code-based simulation tools some metrics are used such as the number of successful attacks, 
potentially  weighted by the total number of attack or the attack surface \cite{Dureuil/CARDIS15}.
Nevertheless these metrics are not really used in practice
to compare robustness of applications: the try and test approach turns out to be suffisant when single fault is considered.
On the contrary, in the context of multi-faults  we have to formalize how implementations can be compared,
that we propose here. Furthermore we will use our formalization to validate our methodology 
and algorithms showing that we improve the robustness of applications in adding countermeasures.

\subsection{Definition of multi-fault robustness}
\label{sec:rob}

Generally simulation tools give the number of attacks for a given \textit{golden run} starting from a fixed initial state. A simulated
trace $t$ is a  finite sequence of transitions corresponding to a nominal execution steps or faulted steps, starting from an initial state.
In the following  $init(t)$ denotes the initial state of $t$ and $fault(t)$ the number of faulted transitions into $t$.  
For a success condition $S$ (generally the negation of a security property), 
a set of fault models $m$, simulated traces $T$ of a program $P$  can be partitioned into the following way:

\vspace{-0.3cm}
\begin{itemize}
    \item \textbf{$T_{N}(m, S)$} : traces obtained under the nominal execution (without fault) 
    \item \textbf{$T_{D}(m, S)$} : faulted traces that are detected by a countermeasure
    \item \textbf{$T_{S}(m, S)$} : successful attacks (verifying $S$ and not detected)
    \item \textbf{$T_{F}(m, S)$} : non detected faulted traces that do not verify  $S$
\end{itemize}

\begin{definition}{Input vulnerability characterization}
\label{def:inputR}

    Let $P$ be a program,   $i$ an  initial state (including inputs),  $m$ a set of fault models and $S$ a successful condition.
We define as $Vuln_{\mu}(P, i,  m, S)$ the 
attack function  associated to the input $i$  defined  from the rank 0 to $\mu$:
\begin{center}
    $Vuln_{\mu}(P,i,  m, S) {\hat =}   f \in 0..\mu \rightarrow \mathbb{N} \mid\  \forall n \in 0..\mu~  (f(n) = \#\{ t\in T_{S}(m, S) \mid\  init(t)=i \ \land\ fault(t) = n \}$

\end{center}
\end{definition}

Definition \ref{def:inputR} can be extended to a set of initial states $I$.
For instance Table \ref{table:exemple2} 
describes $Vuln_{\mu}(P,I,  m, S)$ with $\mu=8$, $I=\{a2[0..3], a1[0..3], 4) \mid \forall i \in 0..3~ a1[i] {\not =} a2[i] \}$,  $m = \{Test\_inversion\}$ and $S$ be {\tt result==BOOL\_TRUE}.

\begin{definition}{Robustness level}
\label{def:robust}

If $I$ is the set of all initial states
and $Vuln_{\mu}(P, i, m, S)(i) = 0$ for all $i\in\ 1..\mu$ and for all $i \in I$
we say that \emph{$P$ is robust up to $\mu$ faults}.
\end{definition}

\subsection{Robustness comparison}

Here we target to compare a program $P$ with an hardened version of $P$.
For instance we want to state that the secure version of function {\tt byteArrayCompare} is more secure
than the initial version, up to order 8.

\begin{definition}{Robustness comparison}
  \label{def:robustness_metric}

Let $P$ be a program, ${I}$ a set of initial states, $\mu$ the order to consider, $m$  the set of  fault models
and $S$ the successful condition. Let $P'$ be an hardened version of $P$. 
Robustness comparison is defined as follow:
\begin{center}
$P' \geq^{rob}_{\mu, {I}, m, S}~ P \  {\hat =}$
  $\forall i\in\ {I}~\forall n\in\ 1..\mu \sum_{j=1}^{n} Vuln_{\mu}(P',i, m, S)(j) \leq \sum_{j=1}^{n} Vuln_{\mu}(P, i, m, S)(j) $
\end{center}
\end{definition}

\noindent
Our definition does not only compare the total number of attacks for a given order but all intermediary 
sums. Thanks to this definition,  several nice properties hold
as {\it monotonicity} ($P' \geq^{rob}_{n+1, {I}, m, S} \ P \ \Rightarrow P'  \geq^{rob}_{n, {I}, m, S} \ P$
)
and transitivity ($P'' \geq^{rob}_{\mu,{I}, m, S} P' \ \land \ P' \geq^{rob}_{\mu, {I}, m, S} P \ \Rightarrow P'' \geq^{rob}_{\mu, {I}, m, S} P$).

\subsection{Discussion} 
A classical approach to evaluate the robustness of a code against fault attack is to count the number of faults issued from a given golden run,
where both the inputs and the initial states are fixed in advance. 
Several works \cite{theissing2013comprehensive,Moro/HOST14,DureuilPPLCC16} use \textit{metrics} to compare protected program versions in this 
setting.  The robustness comparison relation we propose can be seen as a generalization of these metrics, taking into account a {set} of 
initial states as well as multi-faults and still offering good properties.
Note that this comparison relation is relevant only for two programs having the same nominal behaviors (i.e. the same functionality without fault).
Moreover, it is a partial relation. Typically, two programs $P$ and $P'$ can become incomparable up to $n+1$ faults if $P'$ is more robust than $P$ 
up to $n$ faults and, for $n+1$ faults, the sum of the attacks of program $P'$ is greater than the one of $P$. 
Finally, thanks to our symbolic execution based approach,  computing this relation needs to consider in practice only one representative 
attack per execution path (even potentially reduced to  \textit{minimal attacks} as defined in definition~\ref{def:minimal}) hence allowing to reduce metrics explosion 
due to the added countermeasures (according to the dilution paradox~\cite{Dureuil/CARDIS15}).

%% file: cm_analysis.tex
In this section, we propose a systematic approach for analyzing countermeasures " in isolation "
in terms of classes of attacks they detect.
Combined with their proper attack surface (section \ref{sec:pl}) we are able   to construct what we called the "countermeasure catalog", 
used later to choose adapted countermeasures for hardening implementations.
Our approach is based on the encoding of several  generic schemes, described by C codes,
 associated with their pre and postconditions. 
We use  the weakest-precondition WP plugin of  Frama-C and the ACSL annotation language \cite{Kirchner/FAC15}.

\begin{enumerate}
\item  {\bf Mutation schemes} will characterize a fault model as how a fault impacts the  behavior
of sensible  instructions.
Postcondition  states the nominal behavior of the instruction as well as the behavior is impacted by a  fault.

\item {\bf Countermeasure schemes} will describe how abnormal behavior will be detected. 
Postcondition characterized behaviors that are not blocked.

\item {\bf Protected schemes} will describe how countermeasures are inserted w.r.t.
mutation schemes.
The expected postcondition is the nominal behavior associated to the mutation scheme (all faults
provoking a variation of the nominal behavior must be blocked).  
\end{enumerate}

We will illustrate our approach on three fault models: test inversion,  data load modification  (when we read a memory value)
and Else-following-Then. 
These fault models are very classical ones, in particular the Else-following-Then model  describes 
when the jump at the end of the then block is skipped (or transformed into a nop operation).
We also consider three countermeasures: test duplication, load duplication and a lightweight control flow integrity
based on block signature.

\subsection {Mutation schemes}
\label{sec:FMCM}

We consider a fault model as a set of mutation schemes describing how the code is transformed under a fault injection. 
The  {\tt fault} parameter represent the presence of a fault (0 if no fault, different from 0 otherwise).
The postcondition describes the expected nominal behavior (when {\tt fault} equals  0) as well as the result of
a fault. 
Thanks to the ACSL language we can clearly distinguish nominal and faulty behaviors.
Listing \ref{lst:ti_mutation_scheme} describes the mutation scheme for test inversion, when 
inverting the result of a conditional jump. Other fault models are given in Appendix \ref{annexe1}.

\begin{lstlisting}[language=C, caption=Test inversion mutation scheme, label={lst:ti_mutation_scheme}]
/*@ assigns \nothing ; 
  @ behavior nominal :
  @    assumes  fault==0 ;
  @    ensures  C!=0 ==>  \result==br_then ;
  @    ensures  C==0 ==>  \result==br_else ;
  @ behavior faulted :
  @    assumes  fault!=0 ;
  @    ensures  C!=0 ==>  \result==br_else ;
  @    ensures  C==0 ==>  \result==br_then ; @*/
int mutation_ti(int C, int br_then, int br_else, int fault)
   { if((C && !fault) || (!C && fault)){ return br_then; }
     else { return br_else; }  } 
\end{lstlisting}

All fault schemes (listings \ref{lst:ti_mutation_scheme} and \ref{lst:ld_mutation_scheme}, 
\ref{lst:TfE_mutation_scheme} in Appendix \ref{annexe1}) are proved by the plugin Frama-C/WP.

\subsection{Countermeasure schemes}
\label{subsec:cm-schemes}
Countermeasures calls a detection function  (atk\_detected) which stops the execution. 
Listing \ref{lst:testdup_cm} describes the detection function and a countermeasure duplicating 
the condition. Listing \ref{lst:signature_cm}
describes a more sophisticated countermeasure, with a proper variable representing which block must be reached.
Function {\tt sign} records the name of the next block and function {\tt check} verifies that the reached block
corresponds to the expected one.

\begin{lstlisting}[language=C, caption=Test duplication countermeasure, label={lst:testdup_cm}]
/*@  ensures \false; 
  @ assigns \nothing ;@*/
void atk_detected() { exit(0); }

/*@ ensures   C!=0; 
  @ assigns \nothing ; @*/
void TestDup (int C) { if(!C) atk_detected(); }
\end{lstlisting}

\begin{lstlisting}[language=C, caption=Signature based countermeasure, label={lst:signature_cm}]
int RTS;

/*@ ensures  ((C!=0 ==>  RTS==id_true)) ;
  @ ensures  ((C==0 ==>  RTS==id_false)) ;
  @ assigns RTS; @*/
void sign(int C, int id_true, int id_false)
{ if(C) RTS = id_true; else  RTS = id_false; }

/*@ ensures (RTS==id) ;
  @ assigns \nothing; @*/
void check(int id) { if(RTS!=id) atk_detected(); }

\end{lstlisting}

Function {\tt atk\_detected} never normally terminated. Each countermeasure
targets to block an erroneous control flow. 
The postcondition describes the final state ensured by the countermeasure.
All these functions are proved by the WP/Frama-C plugin, as well as the load duplication
countermeasure given in Appendix \ref{annexe1}. 

\subsection{Protected schemes}
\label{sec:protected}

The  last step consists to prove the adequacy of a  countermeasure  against  a fault model. 
To do that we combine countermeasures with
a mutation scheme targeting to block abnormal behaviors introduced by the mutation. 
Listing \ref{insertion_td_on_ti} shows how a condition 
can be protected by test duplication against test inversion.  
The postcondition is build in a systematic way stating that we terminate normally with 
the nominal behavior of the mutation scheme (the nominal behavior of Listing 
\ref{lst:ti_mutation_scheme} for our example below). That means that no fault occurs or a faults occurs but
without impact on the nominal behavior.

\begin{lstlisting}[language=C, caption=The insertion of the TD CM in the TI mutation scheme, label={insertion_td_on_ti}]
/*@ behavior nominal :
  @    ensures  ((C!=0 ==>  \result==br_then)) ;
  @    ensures  ((C==0 ==>  \result==br_else)) ;
  @    assigns \nothing ; @*/

int protected_TI_DT(int C, int br_then, int br_else, int fault){
    if((C && !fault) || (!C && fault))
      { TestDup(!C); return br_then; }
    else { TestDup(C); return br_else; } }

\end{lstlisting}


Listings in Appendix \ref{annexe1} section \ref{sec:protectedCode}
respectively give protected versions for the load modification protected by load duplication 
(listing \ref{lst:ld_cm_scheme}) 
and for the test inversion and the  Else-following-Then fault models protected by the Signature countermeasure 
(listings \ref{lst:TI_sign} and {\ref{lst:TfE_sign}).
All these protected versions (including listing \ref{insertion_td_on_ti}) are proved by the WP/Frama-C plugin.

Contrary to \cite{MartinKP22} we do not prove the reduced postcondition {\tt fault==0} but we impose to preserve the 
nominal  behavior of the instruction impacted by a fault. Our approach is more general: for instance we admit 
a fault that does not impact the nominal behavior, as an attack by test inversion when the two targeted blocks 
are identical or  the data load modification when we inject the right value. 
On the contrary, for listing \ref{lst:TfE_Tdup}, we can strengthen the postcondition with the clause {\tt ensures fault==0}.

\subsection{Results and discussions}

\begin{proposition}{Adequate countermeasure.}
\label{def:adequacy}

Let $m$  be a fault model characterized by a  mutation scheme $MS$ with a nominal
behavior NB.
A protected scheme PS (embedding a countermeasure) is adequate for $m$ if and only if
PS verifies the  postcondition $ NB $.
Then, according to definition \ref{def:robust}  we can state that PS is  robust up to order 1,
for the fault model $m$, for all $I$ verifying the precondition of PS and for the success condition  $\lnot NB$.
\end{proposition}

\begin{table}[H]
\begin{scriptsize}
\begin{center}
\caption{Adequacy of countermeasures against fault models}
\label{table:adequacy}
\begin{tabular}{|l|c|c|c|}
\hline
\backslashbox[8em]{ Fault model}{Countermeasure } & Test duplication & Load duplication & Block signature \\
\hline
Test inversion & adequate & - & adequate \\ 
\hline
Else  following Then & KO & - & adequate \\ 
\hline
Data load modification & - & adequate & - \\
\hline
\end{tabular}
\end{center}
\end{scriptsize}
\end{table}

\vspace*{-1em}

Table \ref{table:adequacy} describes the result of proofs of  protected programs we developed.
They are all proved by 
the WP/FRama-C plugin (adapted) except Listing \ref{lst:TfE_Tdup} that corresponds to an erroneous countermeasure (KO). 
An empty case means that it does not make sense to consider this combination of fault model and
countermeasure.
Due to the modular proofs in WP/Frama-C  we can replace a sensitive instruction
(as a test, a load or and then-else structure) by one of its protected versions without modifying its
nominal behavior.
For instance we can replace the instruction \lstinline$if(C)  return br_then;  else return br_else;$ 
by \lstinline$if(C) {TestDup(!C); return br_then;} else {TestDup(C); return br_else;}$.

To conclude, the notion of adequacy introduced in Proposition~\ref{def:adequacy} tells us whether a countermeasure protects a sensitive code pattern 
against a single-fault attack model.
The proposed methodology is based on the  preservation of the  nominal behavior, that can be carefully specified (the semantics of the code pattern
as well as the {\tt assigns} clause that allows to embed our protected scheme in a larger code without unpredictable side effects).
On the contrary the impact of a fault in mutation scheme is not really useful here but can be exploited later if we want to establish some properties in
the presence of faults.

%% file: algorithms.tex
\input{surface_attack.tex}

\subsection{Our methodology to harden programs}

The approach we propose allows to harden a program $P$ up to $n$ faults, given as a set of fault models $M$ and a 
success condition $S$.
It relies on a \textit{catalog} $C$ of available countermeasures
associated to the model for which they are  \textit{adequate}  and  their  \textit{protection} levels with respect to
the  fault models in $M$.
This approach is based on the following steps:
\begin{enumerate}
	\item Run Lazart to get the set of faults attacks $T_S(M, S)$ up to $n$.  These attacks characterize the robustness of $P$ with respect to $M$ and $S$.
	\item Compute the subset $A$  of minimal attacks (definition \ref{def:minimal})
		and the set $IP$ of their associated injection points (section~\ref{sec:IP}).
		$IP$ gives a superset of the injection points to be protected 
		in order to harden $P$.
	\item Select both a subset $IP' \subseteq IP$ of injection points to protect, together with the corresponding 
		countermeasures available in $C$. 
	\item Build a hardened program $P'$ obtained by protecting each element of $IP'$ using the results of step 2.
\end{enumerate}

The question is now to decide whether $P'$ is robust \textit{by construction} up to $n$ faults.
Two sufficient conditions are  given below (propositions~\ref{prop:single-ip} and \ref{prop:minimal})
leading to the algorithm proposed in Section~\ref{sec:placement-h}.

\begin{proposition}
\label{prop:single-ip}
Let $A$ a set of attacks obtained on a program $P$ with respect to fault models $M$ and a security objective $S$.
Let $P'$ obtained from $P$ by protecting a \textbf{single} $ip$ of each attack trace $a$ of $A$ 
with an adequate countermeasure of protection level $pl>n$. Then $P'$ contains no longer any attack of $A$.
\end{proposition}

To prove this proposition let us consider a $k$-faults attack $a = [ip_1, ip_2, \cdots, ip_k]$ of $A$.
According to Section~\ref{sec:pl}, faulting $ip_i$ in $P'$ requires now $pl+1$ faults, hence each attack of $A$ does no longer succeed in less than $n$ faults. 
Note that this proposition (trivially) holds when the countermeasures cannot be attacked (e.g, they are implemented using tamper-resistant hardware protections), 
since their protection levels can be considered as infinite. 

\begin{proposition}
\label{prop:minimal}
	Thanks to Definition~\ref{def:minimal} on minimal attacks, if we protect every minimal attacks up to $n$, all attacks in $T_S(M, S)$ will be
protected up to $n$. 
\end{proposition}

\subsection{Countermeasure placement algorithm}
\label{sec:placement-h}

Algorithm~\ref{alg:heuristic} aims to protect one $ip$ per execution trace using a countermeasure with a maximal available protection level.  
It iterates over the attacks of $A$ by increasing order of faults injected. 
For each attack $a$, it builds first the set $IP$ of injection points of $a$ with an adequate countermeasure $c$ with a minimal $l$ greater than $n$. 
Then, it selects a candidate $ip$ of $IP$ with the most occurrences in the set of remaining attacks.
If $pl$ is equal to $n$ then attack $a$ becomes protected, together with a maximal number of other attacks sharing this injection point.
Otherwise $a$ is left unprotected.
We denote by $(c,l) \in C(ip)$ a countermeasure and protection level of $C$ of fault model related to the injection point $ip$
(encoded into IP representation as exemplified in Section \ref{sec:IP}).

\begin{algorithm}[ht]
\caption{Heuristic-based hardening algorithm}\label{alg:heuristic}
\begin{algorithmic}
\Require a program $P$, a set of attack traces $A$ of $P$ up to a $n$ faults, a catalog $C$
\State $Protected \gets \emptyset$ \Comment{set of protected traces of $A$}
\State $IpProtected \gets \emptyset$ \Comment{set of protected injection points of $A$}
\For{k in $1$ to $n$}
\State $A_k \gets$ attacks of $A$ with $k$ faults
\For{$a$ in $A_k \setminus Protected$}  \Comment{for all not yet protected attacks}
	\State $IP \gets \{ip \in a \mid \exists (c,l) \in C(ip)~\mbox{s.t. $c$ is adequate for $ip$ and}~ l \geq n\}$ 
\If{$IP \neq \emptyset$} \Comment{keep the ips of $IP$ with a minimal protection level $\geq n$}
	\State $IP \gets \{ip \in IP \mid \exists (c,l) \in C(ip)~\mbox{s.t.}~\forall ip' \in a.\; \forall (c',l') \in C(ip').\; l'>l\}$
	\Else \Comment{take the ips of $a$ with a maximal protection level}
	\State $IP \gets \{ip \in a \mid \exists (c,l) \in C(ip)~\mbox{s.t.}~\forall ip' \in a.\; \forall (c',l') \in C(ip').\; l'<l\}$
\EndIf
\State choose $ip \in IP$ with a maximal number of occurrences in $A \setminus Protected$
\If{$pl+1 > n$}
	\State $IpProtected \gets IpProtected \cup \{(ip,c)\}$ \Comment{$ip$ will be protected with $c$}
	\State $Protected \gets Protected \cup \{a' \in A \mid ip \in a'\}$
\EndIf
\EndFor
\EndFor
\State $P' \gets P$
\For{all $(ip,c)$ in $IpProtected$}
\State $P' \gets$ protect $ip$ of $P'$ with $c$ 
\EndFor
\State return $(P', Protected)$
\end{algorithmic}
\end{algorithm}

\vspace*{-1em}

\subsection{Formal results}
\label{sec:proof}

We can now state results on Algorithm~\ref{alg:heuristic}
based on  Propositions~\ref{prop:single-ip} and \ref{prop:minimal}.

\begin{proposition}{Assurances supplied by algorithm \ref{alg:heuristic}}
\label{prop:algo1}
\begin{enumerate}
\item 
Algorithm\ref{alg:heuristic} terminates.
\item We have $P' \geq^{rob}_{n, {I}, m, S}~ P$ (attacks in $Protected$ or with a prefix in $Protected$ are  no longer successful
attack up to $n$ and no new attack is introduced) 
\end{enumerate}
\end{proposition}

Leveraging the result obtained from $A$ to $P$ needs $A$ to be \textit{representative} of $P$
(i.e., each possible concrete attack of $P$ is represented by a path in the set $T_S(M,S)$).
In our case this means that the path enumeration performed by Lazart to produce $A$ is complete.
Under this hypothesis, if $Protected=A$, then $P'$ is robust up to $n$ faults.


{\bf A degraded version.} If, for some $ip$, $C$ does not contain an adequate countermeasure with a protection level $pl$ greater than $n-1$ 
proposition~\ref{prop:algo1} still holds when replacing $n$ by $x$ faults, where $x$ is 
the minimal protection level used to protect traces of \mbox{$A \setminus Protected$}.

In the general case we cannot say anything beyond $x$ faults, 
because injecting more than $x$ faults on a countermeasure may lead to new execution states and hence new attacks in $P'$.
In this case $P'$ should be hardened as well by repeating the same procedure.
This iterative process terminates either if we get rid of all attacks with strictly less than $n$ faults, or if some $k$-faults attacks (for $k\leq n$)
will always remain unprotected due to a lack of available countermeasure in $C$. 
Furthermore, for particular fault models (as test inversion and data load modification by any value) 
we could be able to show that fault injections  on the protected scheme does not introduce new paths (showing that successful
attacks for the level of protection does not produce states that are not covered by  the associated mutation scheme).
Then, if $A$ is representative of $P$, deviant paths will be already  analyzed\footnote{It is not the case for instance if our data load modification
is reduced to increment the value which is read.}.

%% file: surface_attack.tex
Starting from a set of attacks, the aim is to propose a methodology for assisting countermeasures placement in the more appropriate way.
In the context of multi-faults, countermeasures come with their proper attack surface (the code which is added), and fault injections
on this code  can produce new execution traces, potentially producing new successful attacks.
Because our objective is to propose an alternative to the unrealistic brute force "try and test" approach when multi-faults is considered
we want to study the weakness of countermeasures in a modular way. The next section explains how this analysis can be conducted.

\subsection{Countermeasure protection level} 
\label{sec:pl}

We define (and compute)  the \textbf{protection level} of a countermeasure, which characterizes
the minimal number of faults allowing to bypass a countermeasure.
Bypassing a countermeasure means  the countermeasure terminates normally (without call to the function {\tt atk\_detect})
but violating its postcondition (i.e.  states that must be normally blocked). 

\begin{definition}{Protection level}
\label{def:pl}

\noindent
Let $CM$ be a countermeasure scheme defined by its postcondition $R$ and $M$ a set of fault models.
Let $pl(CM, M)=l$  the level of protection of $CM$ w.r.t. $M$ defined as:

\noindent
{\it \bf Condition 1.} 
l  is the minimal number of necessary faults producing an attack violating $R$:
$l = min \{(fault(t) ~|~ t \in T_S(M, \lnot R)\}$ (or $\infty$ if no successful attack exists).

\noindent
{\it \bf Condition 2.} 
any traces t such that 0<fault(t)<l are blocked :
$\forall t~ (0<fault(t)<l \Rightarrow t \in T_D(M, \lnot R))$
\end{definition}
Sets $T_S(M, \lnot R)$ and $T_D(M, \lnot R)$, introduced section \ref{sec:metrics-analysis},  respectively represent the set of
successful attacks w.r.t. $\lnot R$ and the set of detected attacks (with a call to function {\tt atk\_detected}).
Our tool Lazart allows us to establish the protection level. 
First we compute $T_S(M, \lnot R)$ starting from order 1 to a given bound  $max$ until an attack is found. 
On the contrary Lazart does not directly compute $T_D(M, \lnot R)$.
Then we will  compute $T_S(M, R)$, the set of attacks verifying $R$
 and not detected, that must be empty for 1 to $l-1$ faults (implying that all attacks are detected).
Table \ref{fig:results-k} shows the protection level of our different countermeasures (listings \ref{lst:testdup_cm},  
\ref{lst:signature_cm}, \ref{lst:ld_cm_scheme}) for the two  fault models test inversion and  data load modification and their combination (denoted Comb1 here).

\begin{table}[]
                \caption{Protection levels of countermeasure schemes\label{fig:results-k}}
                \begin{center}
\begin{scriptsize}
\begin{tabular}{l|c|c|c}
\hline
\backslashbox[9em]{Countermeasure}{Fault model } & Test inversion  & Load modification  & Comb1  \\
\hline
Test duplication & 1 & 1 & 1 \\
Load duplication & 1 & 1 & 1 \\
Block signature & 1 & 1 & 1 \\
\hline
$i$ Test duplication  & $i$ & 1 & $i$ \\
$i$ Load duplication   & $i$ & $i$ & $i$ \\
$i$ Block signature   & $i$ & $i$ & $i$ \\
\hline
\end{tabular}
\end{scriptsize}
\end{center}
\end{table}

\vspace*{-1em}

Because all single fault models produce a protection level of 1 it is not mandatory to compute the protection level for the combination
(combinations of models can sometimes allow to produce more efficient successful attacks, but it is not the case for 1).
We extend our protection level when $i$ new instances of countermeasures are considered, because duplicating countermeasures 
preserves the adequacy of protection scheme\footnote{It is also possible to redefined new protection schemes combining or duplicating 
countermeasures and prove their adequacy as before.}.


{\bf Discussion/limitation:} Protection level is a partial function: it is not always possible to compute a protection level when the second condition 
does not hold.  That means that this countermeasure
against the considered fault models can not be analyzed in a modular way. For instance if we consider a fault model jumping anywhere in the code
our countermeasures do not fulfill this condition. In this case countermeasures must embed local verification (for instance on the the value of  
the program counter). 

%% file: experimentations.tex
\subsection{Experimentations}
\label{sec:experimentation}

In this section, we give the results obtained with various countermeasure placement algorithms.
We consider the test inversion and data load mutation fault models}, with their combination, and the 
$i$ Test duplication and $i$ Load duplication countermeasures. Two programs are used for
experimentations:
    \textit{VP} the version 4 of the verifypin program in the FISSC benchmark \cite{DureuilPPLCC16}
    and \textit{FU} an implementation of a firmware updater.
\textit{VP} is a generalization of our  introductive examples (section \ref{sec:motivating-example}) and all these
programs are already be used for large experimentations~\cite{FDTC20} and are freely available\footnote{lazart.gricad-pages.univ-grenoble-alpes.fr/home/fdtc20/index.html}.

Our results are summarized in Table~\ref{tbl:results}.
Columns $PM$  gives the program to protect and the fault models to consider. 
Column IPs gives the total number of injection points of $P$ with respect to $m$.
We consider three  placement algorithms:
    \begin{itemize}
		\item $Naive$: protection of level $n$ for all injection points of $P$.
        \item $All$: protection of level $n$ for all injection points involved in at least one successful attack
		\item $Single$: protection of a single injection point at level $n$ per sucessful attack (Algorithm~\ref{alg:heuristic}).
    \end{itemize}
Column ``\#added-cms'' indicates the number of countermeasures added, for each algorithm, up to $n=4$ faults. 
As expected, Algorithm~\ref{alg:heuristic} gives much better results than the two other ones. In particular the $Naive$ one, used by all tools adding countemeasures in a systematic way 
(disregarding actual attacks) is clearly outperformed. Moreover, Algorithm~\ref{alg:heuristic} significatively reduces the number of added countermeasures, in particular on the FU 
example\footnote{still considering a relevant security property}. 
    
    \begin{table}[ht]
                {\scriptsize
                \begin{center}
\caption{Countermeasures added by placement algorithms\label{tbl:results}}
\begin{tabular}{ll|l|l|llll}
\multicolumn{2}{c|}{} & \multicolumn{1}{c|}{algorithm} & \multicolumn{4}{c}{\#added-cms} \\
PM  & IPs &  & 1 faults & 2 faults & 3 faults & 4 faults   \\
\hline
\hline
VP   & 8 & Naive & 8 & 16 & 24 & 32 \\
with test inversion     &  & All & 3 & 8 & 12 & 16 \\
   &  & Single & 3 & 6 & 9 & 12 \\
 \hline
 \hline
FU  & 42 & Naive & 42 & 84 & 126 & 168 \\
 with Test inversion   &  & All & 0 & 28 & 42 & 88 \\
   &  & Single & 0 & 14 & 21 & 28 \\
 \hline
FU   & 2 & Naive & 2 & 4 & 6 & 8 \\
 with data load mutation   &  & All & 1 & 4 & 6 & 8 \\
   &  & Single & 1 & 2 & 3 & 4 \\
 \hline
FU  &  44 & Naive & 44 & 88 & 132 & 176 \\
 with test inversion  &  & All & 1 & 32 & 60 & 96 \\
   + data load mutation &  & Single & 1 & 16 & 24 & 32 \\
\end{tabular}
                \end{center} 
                }
\end{table}

\vspace*{-1em}




%% file: conclusion.tex
\section{Conclusion}
\label{sec:conclusion}

In this paper we propose a global methodology assisting developer
for hardening applications against multi-fault attacks,
which is nowadays the state of the art \cite{wistp/KimQ07,fdtc/TrichinaK10}.
Generalizing the existing single fault hardening approach is far from easy due to the
he attack surface introduced by countermeasures, potentially introducing 
new attacks and new paths to explore. In order to master this inherent complexity
we propose an innovative approach based on two properties:
the {\it adequacy} of a countermeasure against a given fault model and the {\it protection level},
characterizing it weakness against a set of fault models (number of faults greater than 
the protection level could produce out of control behaviors). The proposed approach is limited 
to countermeasures with a local detection power, as for all automatic tools adding
countermeasures in the single fault context. 
 
At first we give a formal framework for specifying and proving fault models, countermeasures and
protected schemes to establish the adequacy property. 
The key is the specification of the nominal behavior of instructions, how it can be impacted by faults 
and how deviations can be detected.
Our work can be seen as a generalization of the approach described in~\cite{MartinKP22},
also based on the Frama-C/WP plugin, as explained in Section~\ref{sec:protected}. 
Formally establishing the robustness of a countermeasure scheme against an attacker model is not new. In the context
of Control flow integrity many countermeasures have been proposed and proved, as in~\cite{abadi09}.
In the context of fault injection, formal methods have been used 
to establish the effectiveness of countermeasures \cite{Goubet/CARDIS15,heydemann2019formally}.
But these works are dedicated to particular form of countermeasures and specific fault models, and they address single faults only. 

Based on the new notion of {\it protection level} we propose algorithms allowing to
automatically insert countermeasures, going beyond automatic algorithms not taking into consideration
successful attacks (the naive approach). Furthermore we are able to characterize the level of assurance of
the protected programs (property \ref{prop:algo1}). Nevertheless, computing which injection points must be protected, w.r.t.   successful 
properties requires a first analysis giving some exhaustive insurances.  Here we based
this analysis on an symbolic execution engine (Klee \cite{CadarDE08} and Lazart \cite{Potet/ICST14}), which 
is sensitive to the code complexity (number of paths and number of fault injections).
We are currently exploring several ways for mastering this complexity based on static analysis 
\cite{Lacombe21} and a smart encoding of faults \cite{ducousso23}, which could be combined with the approach proposed here.

As pointed out, a such generic framework is new and must be more largely evaluated,  
considering existing fault models and countermeasures, together with 
more sophisticated countermeasures and more flexible placement algorithms. In particular
we are studying an algorithm, based on Integer Linear  Programming,  allowing to combine  several countermeasures per attacks, 
with lower protection levels, still preserving  our robustness properties.

%% file: annexeA.tex
\subsection {Mutation schemes}

\subsubsection {Data load mutation.}
We can mutate a load with the right value, introducing in that a fault with no dangerous impact.

\begin{lstlisting}[language=C,basicstyle=\ttfamily\footnotesize, caption=Data load  mutation scheme, label={lst:ld_mutation_scheme}]
/*@ ensures  assigns \nothing ;@*/
int choose (int value);
/*@ assigns \nothing ;
  @ behavior nominal :
  @    assumes  fault==0 ;
  @    ensures  \result==value;
  @  behavior faulted :
  @    assumes  fault!=0 ;
  @    ensures  \true ; @*/
int mutation_scheme_dl(int value, int fault)
   { int x = value; if(fault) x = choose(value); return x; }
\end{lstlisting}

\subsubsection {Else  following Then mutation.} We encode the flow by
0x10 when we cross the block Then, by 0x01 when we cross the block Else and by 0x11 when we cross the block Then followed by the block Else. 

\begin{lstlisting}[language=C, caption=Else following Then mutation  scheme, label={lst:TfE_mutation_scheme}]
  /*@ behavior nominal :
       @ assumes fault==0  ;
       @ ensures  ((C!=0 ==> \result == 0x10)) ;
       @ ensures  ((C==0 ==>  \result == 0x01)) ;
    @ behavior faulted :
       @ assumes fault!=0  ;
       @ ensures  ((C!=0 ==> \result == 0x11)) ;
       @ ensures  ((C==0 ==> \result == 0x01)) ;@*/
  int mutation_scheme_then_else(int C,  int fault)
  { int FLOW = 0x00;
    if(C) goto then_br;  else goto else_br;
    then_br : FLOW|=0x10; if(!fault) goto fin; 
    else_br : FLOW|=0x01;
    fin : return FLOW; }
\end{lstlisting}

\subsection{Countermeasure schemes}


\begin{lstlisting}[language=C, caption=Load  duplication countermeasure scheme, label={lst:ld_cm_scheme}]
/*@  ensures \result==value;
  @ assigns \nothing ; @*/
void LoadDup(int res, int value)
   {int cm_var  = value; if(res != cm_var) atk_detected();}
\end{lstlisting}

\subsection{Protected schemes}
\label{sec:protectedCode}

\begin{lstlisting}[language=C, caption=Load modification  protected by load duplication, label={ld_cm_scheme}]
/*@ behavior nominal :
  @ ensures \result==value; @*/
int protected_DL_LD(int value, int fault){
    int res = mutation_scheme_dl(value,fault);
    LoadDup(res,value); return(res); }
\end{lstlisting}


\begin{lstlisting}[language=C, caption=TI protected by Signature, label={lst:TI_sign}]
/*@ assigns RTS;
  @ behavior nominal :
  @    ensures  (C!=0 ==>  \result==br_then) ;
  @    ensures  (C==0 ==>  \result==br_else) ; @*/
int protected_EfT_SIG(int C, int br_then, int br_else, int fault){
    sign(C, br_then, br_else);
    if((C && !fault) || (!C && fault)
        { check(br_then); return br_then; }
    else { check(br_else); return br_else; } }
\end{lstlisting}


\begin{lstlisting}[language=C, caption=Else following Then protected by Signature, label={lst:TfE_sign}]
  /*@  requires id_true!=id_false ;
  @ assigns RTS ;
  @ behavior nominal :
  @     ensures  ((C!=0  ==> \result == 0x10)) ;
  @     ensures  ((C==0  ==>  \result == 0x01)) ;@*/
  int protected_ThenElse_CMSign(int C, int id_true, int id_false, int fault)
  { int FLOW = 0x00;
    sign(C,id_true,id_false);
    if(C) goto then_br; else  goto else_br;
    then_br : check(id_true); FLOW|=0X10; if(!fault) goto fin;
    else_br : {check(id_false); FLOW|=0X01;}
    fin : return FLOW; }
\end{lstlisting}

\begin{lstlisting}[language=C, caption=Else following Then protected by Test duplication, label={lst:TfE_Tdup}]
  /*@ behavior nominal :
  @     ensures  ((C!=0 && fault==0 ==> \result == THEN)) ;
  @     ensures  ((C==0  ==>  \result == ELSE)) ;@*/
  int protected_ThenElse_CM_Testdup(int C, int fault)
  { int FLOW = 0x00;
    if(C) { CM_TestDup(C); goto then_br; } else { CM_TestDup(!C); goto else_br; }
    then_br : FLOW|=0x10; if(!fault) goto fin;
    else_br : FLOW|=0x01;
    fin : return FLOW; }
\end{lstlisting}

%% file: main.bbl
\begin{thebibliography}{10}
\providecommand{\url}[1]{#1}
\csname url@samestyle\endcsname
\providecommand{\newblock}{\relax}
\providecommand{\bibinfo}[2]{#2}
\providecommand{\BIBentrySTDinterwordspacing}{\spaceskip=0pt\relax}
\providecommand{\BIBentryALTinterwordstretchfactor}{4}
\providecommand{\BIBentryALTinterwordspacing}{\spaceskip=\fontdimen2\font plus
\BIBentryALTinterwordstretchfactor\fontdimen3\font minus
  \fontdimen4\font\relax}
\providecommand{\BIBforeignlanguage}[2]{{%
\expandafter\ifx\csname l@#1\endcsname\relax
\typeout{** WARNING: IEEEtran.bst: No hyphenation pattern has been}%
\typeout{** loaded for the language `#1'. Using the pattern for}%
\typeout{** the default language instead.}%
\else
\language=\csname l@#1\endcsname
\fi
#2}}
\providecommand{\BIBdecl}{\relax}
\BIBdecl

\bibitem{wistp/KimQ07}
C.~H. Kim and J.~Quisquater, ``Fault attacks for {CRT} based {RSA:} new
  attacks, new results, and new countermeasures,'' in \emph{Information
  Security Theory and Practices. {WISTP} 2007, Heraklion, Greece, 2007,
  Proceedings}, 2007, pp. 215--228.

\bibitem{fdtc/TrichinaK10}
E.~Trichina and R.~Korkikyan, ``Multi fault laser attacks on protected
  {CRT-RSA},'' in \emph{2010 Workshop on Fault Diagnosis and Tolerance in
  Cryptography, {FDTC} 2010, Santa Barbara, California, USA, 21 August 2010},
  2010, pp. 75--86.

\bibitem{inter_cesti20}
ANSSI, Amossys, EDSI, LETI, Lexfo, Oppida, Quarkslab, SERMA, Synacktiv, Thales,
  and T.~Labs, ``Inter-cesti: Methodological and technical feedbacks on
  hardware devices evaluations,'' in \emph{SSTIC 2020, Symposium sur la
  sécurité des technologies de l'information et des communications}, 2020.

\bibitem{MartinKP22}
T.~Martin, N.~Kosmatov, and V.~Prevosto, ``Verifying redundant-check based
  countermeasures: a case study,'' in \emph{{SAC} '22: The 37th {ACM/SIGAPP}
  Symposium on Applied Computing}.\hskip 1em plus 0.5em minus 0.4em\relax
  {ACM}, 2022.

\bibitem{Seaborn15}
M.~Seaborn and T.~Dullien, ``{E}xploiting the {DRAM} {R}owhammer bug to gain
  kernel privileges: how to cause and exploit single bit errors,'' in
  \emph{Black Hat}, 2015.

\bibitem{GrussMM16}
D.~Gruss, C.~Maurice, and S.~Mangard, ``Rowhammer.js: {A} remote
  software-induced fault attack in javascript,'' in \emph{{DIMVA} 2016, San
  Sebasti{\'{a}}}, 2016, pp. 300--321.

\bibitem{Tang17}
A.~Tang, S.~Sethumadhavan, and S.~J. Stolfo, ``{CLKSCREW:} exposing the perils
  of security-oblivious energy management,'' in \emph{26th {USENIX} Security
  Symposium, {USENIX} Security 2017}.\hskip 1em plus 0.5em minus 0.4em\relax
  {USENIX} Association, 2017.

\bibitem{QiuWLQ19}
P.~Qiu, D.~Wang, Y.~Lyu, and G.~Qu, ``Voltjockey: Breaching trustzone by
  software-controlled voltage manipulation over multi-core frequencies,'' in
  \emph{Proceedings of the 2019 {ACM} {SIGSAC} Conference on Computer and
  Communications Security, {CCS}}.\hskip 1em plus 0.5em minus 0.4em\relax
  {ACM}, 2019.

\bibitem{Murdock2019plundervolt}
K.~Murdock, D.~Oswald, F.~D. Garcia, J.~Van~Bulck, D.~Gruss, and F.~Piessens,
  ``{Plundervolt}: Software-based fault injection attacks against {I}ntel
  {SGX},'' in \emph{{Proceedings of the 41st IEEE Symposium on Security and
  Privacy (S\&P'20)}}, 2020.

\bibitem{Timmers16}
N.~Timmers and A.~Spruyt, ``Bypassing secure boot using fault injection,'' in
  \emph{Black hat Europe 2016}.

\bibitem{app12010417}
S.~Delarea and Y.~Oren, ``Practical, low-cost fault injection attacks on
  personal smart devices,'' \emph{Applied Sciences}, vol.~12, no.~1, 2022.

\bibitem{Hillebold2014}
C.~Hillebold, ``Compiler-assisted integrity against fault injection attacks,''
  Master's thesis, University of Technology, Graz, December 2014.

\bibitem{ProyHBC17}
J.~Proy, K.~Heydemann, A.~Berzati, and A.~Cohen, ``Compiler-assisted loop
  hardening against fault attacks,'' \emph{{ACM} Trans. Archit. Code Optim.},
  vol.~14, no.~4, pp. 36:1--36:25, 2017.

\bibitem{BellevilleHCBRSC18}
N.~Belleville, K.~Heydemann, D.~Courouss{\'{e}}, T.~Barry, B.~Robisson,
  A.~Seriai, and H.~Charles, ``Automatic application of software
  countermeasures against physical attacks,'' in \emph{Cyber-Physical Systems
  Security.}, 2018, pp. 135--155.

\bibitem{Potet/ICST14}
M.-L. Potet, L.~Mounier, M.~Puys, and L.~Dureuil, ``Lazart: {A} symbolic
  approach for evaluation the robustness of secured codes against control flow
  injections,'' in \emph{Seventh {IEEE} International Conference on Software
  Testing, Verification and Validation, {ICST} 2014}.\hskip 1em plus 0.5em
  minus 0.4em\relax IEEE, 2014, pp. 213--222.

\bibitem{DureuilPPLCC16}
L.~Dureuil, G.~Petiot, M.~Potet, T.~Le, A.~Crohen, and P.~de~Choudens,
  ``{FISSC:} {A Fault Injection and Simulation Secure Collection},'' in
  \emph{Computer Safety, Reliability, and Security - 35th International
  Conference, {SAFECOMP} 2016}, 2016, pp. 3--11.

\bibitem{CadarDE08}
C.~Cadar, D.~Dunbar, and D.~R. Engler, ``{KLEE:} unassisted and automatic
  generation of high-coverage tests for complex systems programs,'' in
  \emph{8th {USENIX} Symposium on Operating Systems Design and Implementation,
  {OSDI} 2008, USA}, 2008, pp. 209--224.

\bibitem{FDTC20}
E.~Boespflug, C.~Ene, L.~Mounier, and M.-L. Potet, ``{Countermeasures
  Optimization in Multiple Fault-Injection Context},'' in \emph{{''Fault
  Diagnosis and Tolerance in Cryptography'' FDTC 2020}}, Sep. 2020.

\bibitem{Lacombe21}
G.~Lacombe, D.~Feliot, E.~Boespflug, and M.-L. Potet, ``Combining static
  analysis and dynamic symbolic execution in a toolchain to detect fault
  injection vulnerabilities,'' \emph{Journal of Cryptographic Engineering},
  vol. 18 january 2023.

\bibitem{Christofi/CE13}
M.~Christofi, B.~Chetali, L.~Goubin, and D.~Vigilant, ``{Formal verification of
  a CRT-RSA implementation against fault attacks},'' \emph{Journal of
  Cryptographic Engineering}, vol.~3, no.~3, pp. 157--167, 2013.

\bibitem{Rauzy/CE14}
P.~Rauzy and S.~Guilley, ``\BIBforeignlanguage{English}{{A Formal Proof of
  Countermeasures Against Fault Injection Attacks on CRT-RSA}},''
  \emph{\BIBforeignlanguage{English}{Journal of Cryptographic Engineering}},
  vol.~4, no.~3, pp. 173--185, 2014.

\bibitem{Goubet/CARDIS15}
L.~Goubet, K.~Heydemann, E.~Encrenaz, and R.~De~Keulenaer, ``{E}fficient
  {D}esign and {E}valuation of {C}ountermeasures against {F}ault {A}ttack with
  {F}ormal {V}erification,'' in \emph{14th {S}mart {C}ard {R}esearch and
  {A}dvanced {A}pplication {C}onference, {CARDIS}}, Nov. 2015.

\bibitem{heydemann2019formally}
K.~Heydemann, J.-F. Lalande, and P.~Berthom{\'e}, ``Formally verified software
  countermeasures for control-flow integrity of smart card c code,''
  \emph{Computers \& Security}, vol.~85, pp. 202--224, 2019.

\bibitem{common-criteria-3}
\BIBentryALTinterwordspacing
T.~CCRA Management~Committee. (2012, Sep.) {Common Criteria for Information
  Technology Security Evaluation}. [Online]. Available:
  \url{http://www.commoncriteriaportal.org/files/ccfiles/CCPART3V3.1R4.pdf}
\BIBentrySTDinterwordspacing

\bibitem{JIL/20}
``{Application of Attack Potential to Smartcards and Similar Devices},'' Joint
  Interpretation Library, Tech. Rep. Version 3.0, June 2020.

\bibitem{Dureuil/CARDIS15}
L.~Dureuil, M.-L. Potet, P.~d. Choudens, C.~Dumas, and J.~Cl\'edi\`ere, ``From
  code review to fault injection attacks: Filling the gap using fault model
  inference,'' in \emph{14th Smart Card Research and Advanced Application
  Conference, CARDIS15}.\hskip 1em plus 0.5em minus 0.4em\relax {LNCS}, 2015.

\bibitem{theissing2013comprehensive}
N.~Theissing, D.~Merli, M.~Smola, F.~Stumpf, and G.~Sigl, ``Comprehensive
  analysis of software countermeasures against fault attacks,'' in \emph{2013
  Design, Automation \& Test in Europe Conference \& Exhibition (DATE)}.\hskip
  1em plus 0.5em minus 0.4em\relax IEEE, 2013, pp. 404--409.

\bibitem{Moro/HOST14}
N.~Moro, K.~Heydemann, A.~Dehbaoui, B.~Robisson, and E.~Encrenaz,
  ``Experimental evaluation of two software countermeasures against fault
  attacks,'' in \emph{{IEEE} International Symposium on Hardware-Oriented
  Security and Trust, {HOST} 2014}.

\bibitem{Kirchner/FAC15}
F.~Kirchner, N.~Kosmatov, V.~Prevosto, J.~Signoles, and B.~Yakobowski,
  ``Frama-c: {A} software analysis perspective,'' \emph{Formal Asp. Comput.},
  vol.~27, no.~3, pp. 573--609, 2015.

\bibitem{abadi09}
M.~Abadi, M.~Budiu, {\'U}.~Erlingsson, and J.~Ligatti, ``Control-flow integrity
  principles, implementations, and applications,'' \emph{ACM Transactions on
  Information and System Security (TISSEC)}, vol.~13, no.~1, pp. 1--40, 2009.

\bibitem{ducousso23}
S.~Ducousso, S.~Bardin, and M.-L. Potet, ``Adversarial reachability for
  program-level security analysis,'' in \emph{European Symposium on Programming
  (ESOP)}, 2023.

\end{thebibliography}
